\documentclass[letterpaper]{ptephy}

\usepackage{boites}

\usepackage{mathrsfs}

\begin{document}

\title{
IceCube PeV--EeV neutrinos and secret interactions of neutrinos
}

\author{Kunihito Ioka$^{1}$ and Kohta Murase$^{2}$}

\address{
${}^{1}$Theory Center, Institute of Particle and Nuclear Studies, KEK; 
Department of Particle and Nuclear Physics, 
the Graduate University for Advanced Studies (Sokendai), 
Tsukuba 305-0801, Japan\\
${}^{2}$
Hubble Fellow -- Institute for Advanced Study, Princeton, New Jersey 08540, USA
}

\begin{abstract}%
We show that the PeV neutrinos detected by IceCube 
put unique constraints on ``secret'' interactions of neutrinos
with the cosmic neutrino background (C$\nu$B).
The coupling must be 
$g <0.03$ for the mediating boson mass $m_{X} \lesssim 2$ MeV,
$g/m_{X} < 5$ GeV$^{-1}$ for $m_{X} \gtrsim 20$ MeV,
and $g/m_{X} < 0.07$ GeV$^{-1}$ in between.
We also investigate the possibility that neutrino cascades 
degrade high-energy neutrinos to PeV energies
by upgrading C$\nu$B
where the energy flux of PeV neutrinos can 
coincide with the Waxman--Bahcall bound or 
the cosmogenic neutrino flux for protons, thanks to energy conservation.
However, a large coupling is required, which is disfavored
by laboratory decay constraints.  
The suppression of PeV--EeV neutrinos is a testable prediction
for the Askaryan Radio Array.
\end{abstract}

\subjectindex{E4, F2, B7}

\maketitle

\section{Introduction}

Recently IceCube reported the detection of two PeV neutrinos
and 26 additional events, more than expected from atmospheric backgrounds
\cite{IC13:1stPeV,IC13:Sci}.
The arrival direction is consistent with the isotropic distribution,
suggesting that at least some of the events are of cosmological origin. 
The Hillas condition to accelerate primary cosmic-rays up to $\sim 100$ PeV
allows a dozen possibilities \cite{lah+13,Anchordoqui+13},
such as gamma-ray bursts \cite{Waxman_Bahcall97,Murase+06,gz07,Murase_Ioka13}, 
active galactic nuclei \cite{ste13,Murase+14}, 
galaxy clusters and groups \cite{Murase+08,Murase+13:hadronuclear}, 
star-forming galaxies \cite{Loeb_Waxman06,Murase+13:hadronuclear,kat+13}, 
and heavy dark matter \cite{Feldstein+13,Esmaili_Serpico13,Ema+13}.
We are witnessing the birth of high-energy neutrino astrophysics.

The IceCube events remind us of Supernova (SN) 1987A,
which placed unique limits on the properties of neutrinos,
especially ``secret'' interactions of neutrinos
with the cosmic neutrino background (C$\nu$B) \cite{Kolb_Turner87,Manohar87}.
The neutrino--neutrino interactions \cite{Bialynicka-Birula64,Bardin+70},
even stronger than the weak interactions of the standard model,
remain largely unconstrained below the electroweak energy scale
\cite{Bilenky_Santamaria99,Lessa_Peres07,Laha+13}
because of their weakness and the difficulties in focusing the neutrino beam \cite{T2K14}.
High-energy neutrinos are attenuated by C$\nu$B if the cross section is large enough \cite{Weiler82}.
The much longer distance and higher energy of the IceCube events than those of SN 1987A
can tighten restrictions on the secret interactions \cite{Keranen98},
as well as neutrino decays \cite{bae+12,Pakvasa+13},
leptoquark couplings \cite{Barger_Keung13}, and so on.

The IceCube neutrinos may even result from secret interactions of neutrinos.
By energizing C$\nu$B, high-energy neutrinos may develop cascades
in intergalactic space, 
like gamma-ray cascades
\citep[e.g.,][]{Coppi_Aha97,Murase+12,Inoue_Ioka12}.
Since such self-interactions conserve the total energy,
the neutrinos keep the energy flux while reducing the typical energy.
Thus, this scenario can naturally account for a possible ``{\it coincidence problem}'':
why the observed neutrino flux is comparable to the Waxman--Bahcall bound \cite{Waxman_Bahcall99}
or equivalently the cosmogenic neutrino flux at EeV energies produced by ultrahigh-energy 
cosmic-ray (UHECR) protons \citep[e.g.,][]{bz69,Yoshida_Teshima93,Takami+09}.
The lack of $>2$ PeV events indicates either a soft spectrum or a break 
at several PeV \cite{IC13:Sci,lah+13},
implying different processes at PeV and EeV energies.
It is a coincidence that two different processes
separated by three orders of magnitude in energy
give almost the same flux.

We use 
$(H_0,\Omega_m,\Omega_{\Lambda})
=(72\ {\rm km}\ {\rm s}^{-1}\ {\rm Mpc}^{-1}, 0.27, 0.73)$
and $c=\hbar=k=1$.

\section{Neutrino--neutrino interactions beyond the Standard Model}

We consider non-standard neutrino interactions between themselves, 
through scalar 
${\mathscr L}_{\rm int} = g_{ij} \bar \nu_{i} \nu_{j} \phi$
or pseudoscalar bosons
${\mathscr L}_{\rm int} =g'_{ij} \bar \nu_{i} \gamma^5 \nu_{j} \phi$
as in Majoron-like models \cite{Chikashige+81,Gelmini_Roncadelli81,Georgi+81,Schechter_Valle82},
or vector bosons
${\mathscr L}_{\rm int} =g_{ij} \bar \nu_{i} \gamma^{\mu} \nu_{j} X_{\mu}$
\cite{Kolb_Turner87,Laha+13}.
We assume that a boson has mass 
$m_{X} \sim$ MeV--GeV,
and does not directly
couple (or couples very weakly) to charged particles
to evade experimental constraints.
There exist gauge-invariant models under
electroweak
$SU(2)$ \cite{Choi_Santamaria91,Gavela+09}.

The cross section for scattering $\nu \nu \to \nu \nu$ 
is generally written as~\citep[e.g.,][]{Kolb_Turner87,Goldberg+06}
\begin{eqnarray}
  \sigma_{\nu\nu} \simeq
  \frac{g^4}{16\pi}\frac{s}{(m_X^2-s)^2+m_X^2 \Gamma_X^2}
\simeq \left\{
\begin{array}{ll}
\frac{1}{16\pi} (g^{2}/m_{X}^{2})^2 s, &  (s \ll m_{X}^2)\\
\frac{1}{16\pi} g^{4}/s, &  (s \gg m_{X}^2)
\end{array}\right.
\label{eq:signunu}
\end{eqnarray}
where $\sqrt{s}$ is the center-of-mass energy
and $\Gamma_{X} \simeq g^2 m_{X}/4\pi$ is the decay width.
In the low-energy limit the interaction is described 
by the Fermi's four-fermion theory,
while
in the high-energy limit 
the boson mass is negligible.
At a resonance $s \approx m_{X}^2$, 
we obtain $\sigma_{\nu\nu}\sim\pi/m_X^2$.  
For cosmological sources at $z$, 
a $\delta$-function approximation for the resonance
gives $\sigma_{\nu \nu}^{\rm eff}\sim \pi g^2/(4 m_X^2)$ 
for $\frac{m_X^2}{(1+z)2m_\nu}<\varepsilon_{\nu}^{\rm obs}<\frac{m_X^2}{2m_\nu}$
\cite{Goldberg+06}.
In addition, the annihilation $\nu \nu \to X X \to \nu \nu \nu \nu$ 
contributes
$\sigma_{\nu\nu} \propto (g^4/s) \ln(s/m_{X}^2)$ for $s \gg m_{X}^2$.   
We do not distinguish the types of bosons
nor neutrino--antineutrino.
Our discussion is basically applicable
if a single flavor or a single pair of flavors exchange energy, 
e.g., $\nu_e \nu_{\tau} \to \nu_e \nu_{\tau}$,
because of flavor mixing.

For high-energy neutrinos interacting with C$\nu$B,
the cross section (\ref{eq:signunu})
may be regarded as a function of the energy $\varepsilon_{\nu}$
of the high-energy neutrinos by using the relation
$s \simeq 2 m_{\nu} \varepsilon_{\nu}$,
where we take a neutrino mass $m_{\nu} \sim 0.05$ eV as a fiducial value.
From neutrino oscillations, at least one flavor has mass $m_{\nu} \gtrsim 0.05$ eV
and the other has $m_{\nu} \gtrsim 0.009$ eV,
while the cosmological observations limit
the sum of the masses as $\sum m_{\nu} \lesssim 0.3$ eV
\cite[e.g.,][]{Planck14}.
Note that we should use $T_{\nu}$ instead of $m_{\nu}$ 
if $m_\nu$ is less than the C$\nu$B temperature
$T_{\nu}=(4/11)^{1/3} T_{\gamma} \simeq 1.95$ K $=1.68 \times 10^{-4}$ eV.
For different masses $m'_{\nu}$ (or $T_{\nu}$),
our results can be scaled by
$s \to s (m'_{\nu}/m_{\nu})$,
$g \to g (m'_{\nu}/m_{\nu})^{1/4}$,
and $m_{X} \to m_{X} (m'_{\nu}/m_{\nu})^{1/2}$.

\begin{figure}
  \begin{center}
    \includegraphics[width=8cm]{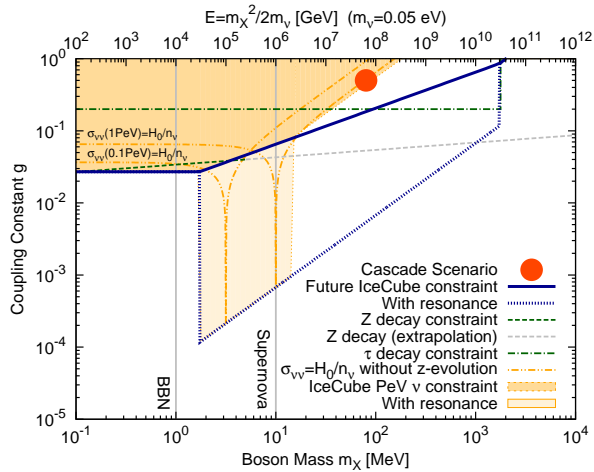}
  \end{center}
  \caption{
    Constraints on the hidden interactions of neutrinos
    with the coupling constant $g$ and the boson mass $m_{X}$,
    with and without resonance
    using the second and the last term in Eq.~(\ref{eq:signunu}),
    respectively.
    The shaded (blue) regions are excluded 
    by the observed (future) IceCube signal
    with Eq.~(\ref{eq:att}).
    We illustrate spiky constraints for a given neutrino energy, 
    although we use $\sigma_{\nu\nu}^{\rm eff}$ for the resonance constraints.
    Also shown are the constraints by the $Z$-decay width
    for the light vector mediator \cite{Laha+13} and
    by the $\tau$-decay rate.
    The left regions of the vertical lines are relevant to
    BBN and supernovae. 
  }
  \label{fig:g}
\end{figure}

The high-energy neutrinos are attenuated if
the mean free path $\lambda_{\nu}=1/n_{\nu} \sigma_{\nu\nu}$
is smaller than the distance to the source $d$,
where $n_{\nu}=\frac{1}{2} \times \frac{3}{11} n_{\gamma} \simeq 56$ cm$^{-3}$ is
the current number density of C$\nu$B for each type ($\nu$ or $\bar \nu$), 
neglecting neutrino asymmetry \cite{Dolgov+02}.
For extragalactic sources at a cosmological distance $d \sim c H_0^{-1}$,
the attenuation condition is
\begin{eqnarray}
\sigma_{\nu\nu}(\varepsilon_{\nu}) > \frac{H_0}{c n_{\nu}}
\sim 1.4 \times 10^{-30}\ {\rm cm}^{2},
\quad (s\simeq2 m_{\nu} \varepsilon_{\nu}).
\label{eq:att}
\end{eqnarray}
With the PeV events in $3\times 10^{4}\ {\rm GeV} \lesssim \varepsilon_{\nu} 
\lesssim 2 \times 10^{6}$ GeV,
we can constrain the coupling $g$ in Eq.~(\ref{eq:signunu}) for each $m_{X}$ 
in Fig.~\ref{fig:g} ({\it shaded}). 
It scales as $g\approx$ const. for $m_X\lesssim2$~MeV, 
$g\propto m_X^{1/2}$ [without resonance 
using the last term in Eq.~(\ref{eq:signunu})]
and $g\propto m_X$ [with resonance 
using the second term in Eq.~(\ref{eq:signunu})]
for 2~MeV~$\lesssim m_X\lesssim 20$~MeV, and 
$g\propto m_X$ for $m_X\gtrsim 20$~MeV.
Further events in 
$2\times 10^{6}\ {\rm GeV} \lesssim \varepsilon_{\nu} \lesssim 3 \times 10^{10}$ GeV
can improve the limits
if detected
({\it blue lines}).
The limits also become more strict
if the source evolution peaks at higher redshifts.

The IceCube limits are independent of the previous constraints 
\cite{Bilenky_Santamaria99,Lessa_Peres07,Laha+13}:
\begin{itemize}
\item {\bf Decay measurements}:
  The decay width of the $Z$ gauge boson
  strongly constrains the hidden neutrino interactions.
  Emission of bosons or neutrinos from a final state neutrino
  or their loop increases the decay width,
  which is precisely measured by experiments.
  We extrapolate the result
  for the light vector mediator
  \cite{Laha+13} as in Fig.~\ref{fig:g}.
  This extrapolation
  would become worse as $m_{X}$ gets closer to the $Z$ mass. 
  In the heavy mass limit, the contact four-fermion neutrino interactions
  cannot be larger than the standard neutral current interactions
  \cite{Bilenky_Santamaria99},
  i.e.,
  $g\sim 0.6$ at $m_{X}\sim 90$ GeV in Fig.~\ref{fig:g}.
  The decay rates of mesons such as $\pi$, $K$, $D$, and $B$
  also provide $g_{ej} < 10^{-2}$, $g_{\mu j} < 0.5$, and $g_{\tau j} < 3$
  \cite{Barger+82,Lessa_Peres07}.
  The decay rates of $\mu$ and $\tau$ also give
  $g_{\mu j} < 10^{-2}$ and $g_{\tau j} < 0.2$ (see Fig.~\ref{fig:g}).

\item {\bf Neutrinoless double beta decay} (0$\nu \beta \beta$):
  The 0$\nu \beta \beta$ bounds $g_{ee}< 10^{-5}$ \cite{Gando+12}.
  
\item{\bf Supernova observations}:
  Neutrino detections from SN 1987A gave $g/m_{X} < 12$ MeV$^{-1}$ \cite{Kolb_Turner87,Manohar87}.
  New bosons could modify the supernova dynamics if $m_{X} \lesssim 10$ MeV.
  In the majoron models, the lepton number is not conserved,
  leading to entropy production and a thermal bounce of the core at a subnuclear density
  \cite{Fuller+88}.
  For a large cross section $\sigma_{\nu\nu}$,
  the neutrino-sphere becomes large, but
  this does not cause too long a delay of the neutrino arrival time
  nor different neutrino heating for explosions,
  because the evolution of the relativistic neutrino gas 
  is largely independent of $\sigma_{\nu\nu}$ \cite{Dicus+89},
  similar to gamma-ray burst fireballs.
  Nevertheless the flavor evolution could be different \cite{Blennow+08}.

\item {\bf Big bang nucleosynthesis (BBN)}:
  If $m_{X}<1$ MeV, the bosons increase
  the effective relativistic degrees of freedom,
  while they may help the small-scale structure problems 
  in the $\Lambda$CDM cosmology \cite{vandenAarssen+12}.
  BBN limits are strong if the right-handed neutrinos
  participate in the interaction \cite{Masso_Toldra94}.



\item {\bf Cosmic microwave background (CMB)}:
  The CMB could provide limits if the neutrino interactions suppress the anisotropic stress
  \cite{Archidiacono_Hannestad13} (but see \cite{Dicus+89}).

\end{itemize}

\section{Neutrino cascades via secret interactions}

The total energy of the neutrinos is conserved in the neutrinophilic interactions because
the boson immediately decays back to neutrinos $X \to \nu \nu$
with a rate $\Gamma_{X} \sim g^2m_X/4\pi$ as long as $m_{X} > 2 m_{\nu}$.
Then 
high-energy neutrinos, in particular EeV cosmogenic neutrinos,
can cascade down to PeV neutrinos 
by upscattering
C$\nu$B.
Interestingly, this scenario could explain why 
the PeV neutrino flux is comparable to the Waxman--Bahcall bound
or the cosmogenic neutrino flux for pure protons.

To calculate the neutrino spectra,
we consider the simplified Boltzmann equations,
\begin{eqnarray}
\!\!\!\frac{\partial f_p(\varepsilon_p^{\rm obs},z)}{\partial t}
\!\!\!
&=&
\!\!\! (1+z) \frac{dn_p}{dt d\varepsilon_p} 
(\varepsilon_p,z)
-\frac{c}{\lambda_p(\varepsilon_p,z)} f_p(\varepsilon_p^{\rm obs},z) 
K_{\pi},
\label{eq:dfpdt}
\\
\!\!\!\frac{\partial f_{\nu}(\varepsilon_{\nu}^{\rm obs},z)}{\partial t}
\!\!\!
&=&
\!\!\! \frac{3c}{2\lambda_p(\varepsilon_{p},z)} f_p(\varepsilon_p^{\rm obs},z)
\frac{\partial \varepsilon_{p}^{\rm obs}}{\partial \varepsilon_{\nu}^{\rm obs}}
- \frac{c}{\lambda_{\nu}(\varepsilon_{\nu},z)} 
f_{\nu}(\varepsilon_{\nu}^{\rm obs},z)
+ \frac{4c}{\lambda_{\nu}(2 \varepsilon_{\nu},z)} 
f_{\nu}(2 \varepsilon_{\nu}^{\rm obs},z),
\label{eq:dfnudt}
\end{eqnarray}
where $f_i(\varepsilon_i,t)$ [cm$^{-3}$ GeV$^{-1}$] 
is the homogeneous and isotropic distribution function (the number of particles per comoving volume per energy)
for UHECR protons ($i=p$) and neutrinos ($i=\nu$),
$\varepsilon_i$ and $t$ are the energy and the proper time 
measured by the comoving observer,
and we have changed the independent variables as
$(\varepsilon_i,t) \to (\varepsilon_i^{\rm obs},z)\equiv (\varepsilon_i/(1+z),z)$
to take the cosmological redshift into account.
Note $|dt/dz|^{-1}=H_0(1+z)\sqrt{\Omega_m(1+z)^3+\Omega_{\Lambda}}$.  

We consider neutrinos from UHECR protons via $p\gamma$ interactions with CMB,
assuming the typical neutrino energy 
$\varepsilon_{\nu} \approx 0.05 \varepsilon_{p}$
and the mean inelasticity $K_{\pi} \sim 0.2$.
The 
mean free paths for protons and neutrinos are given by
\begin{eqnarray}
\frac{1}{\lambda_p(\varepsilon_p,z)}
= \frac{1}{2 \gamma_{p}^2}
\int_{\varepsilon'_{\rm th}}^{\infty} d\varepsilon'_{\gamma} \sigma_{p\gamma}(\varepsilon'_{\gamma})
\varepsilon'_{\gamma}
\int_{\varepsilon'_{\gamma}/2\gamma_p}^{\infty}
d\varepsilon_{\gamma} \frac{1}{\varepsilon_{\gamma}^2}
\frac{dn_{\gamma}}{d\varepsilon_{\gamma}}(\varepsilon_{\gamma},z),
\quad
\frac{1}{\lambda_{\nu}(\varepsilon_{\nu},z)}
= n_{\nu}(z) \sigma_{\nu \nu}(\varepsilon_{\nu}),
\label{eq:lanu}
\end{eqnarray}
where $\gamma_p = \varepsilon_p/m_p c^2$ and we evaluate Eq.~(\ref{eq:lanu}) with the $\Delta$-resonance approximation
$\sigma_{p\gamma}^{\Delta} \sim 4 \times 10^{-28}$ cm$^{-2}$, $\varepsilon'_{\Delta} \sim 0.3$ GeV, and $\Delta \varepsilon' \sim 0.2$ GeV
\cite{Stecker68,Waxman_Bahcall97}.  
For simplicity, we neglect the Bethe--Heitler process. Although the resulting neutrino flux is overestimated by a factor of 2--3, it is enough for our purpose due to uncertainty of redshift evolution models.  Although an approximation $\varepsilon_{\nu} \approx 0.05 \varepsilon_{p}$ also affects the low-energy side of the peak 
of cosmogenic neutrinos by a factor and multi-pion production is relevant at high energies \cite{Takami+09}, we 
may neglect these effects as long as details of the original spectrum are smeared by cascades. 

For simplicity, 
we assume that a high-energy neutrino imparts half of its energy to C$\nu$B
via new interactions.
The influence on details 
of spectra is less than a factor of two.

The energy generation rate of UHECRs is normalized 
by the observations at $10^{19.5}$ eV 
\cite{kat+13}
\begin{eqnarray}
\varepsilon_p^2 \frac{dn_p}{dt d\varepsilon_p}
(\varepsilon_p,z)
\approx 
0.5
\times 10^{44}\ {\rm erg}\ {\rm Mpc}^{-3}\ {\rm yr}^{-1}
(p-1)
\left(\frac{\varepsilon_p}{10^{19.5}\ {\rm eV}}\right)^{2-p}
\exp\left(-\frac{\varepsilon_p}{10^{21}\ {\rm eV}}\right)
R(z),
\label{eq:zevo}
\end{eqnarray}
where we use $p = 2$
and the source evolution 
$R(z)=(1+z)^{4}$ for $z<1.2$
and $R(z) \propto (1+z)^{-1.2}$ for $1.2<z$ for demonstrative purposes
\cite[cf.][]{Takami+09,de Souza+11}.

\begin{figure}
  \begin{center}
    \includegraphics[width=7cm]{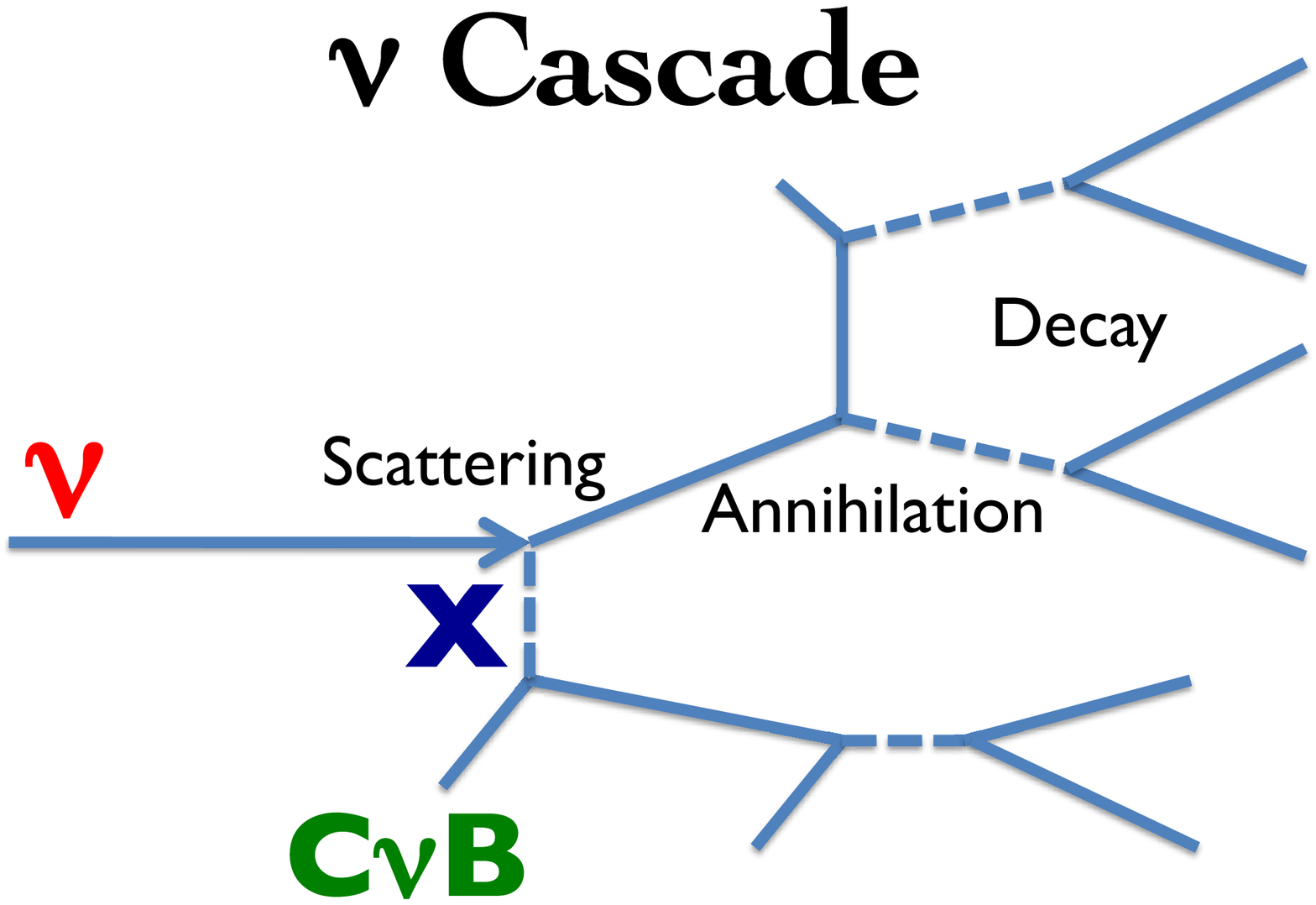}
    \includegraphics[width=8cm]{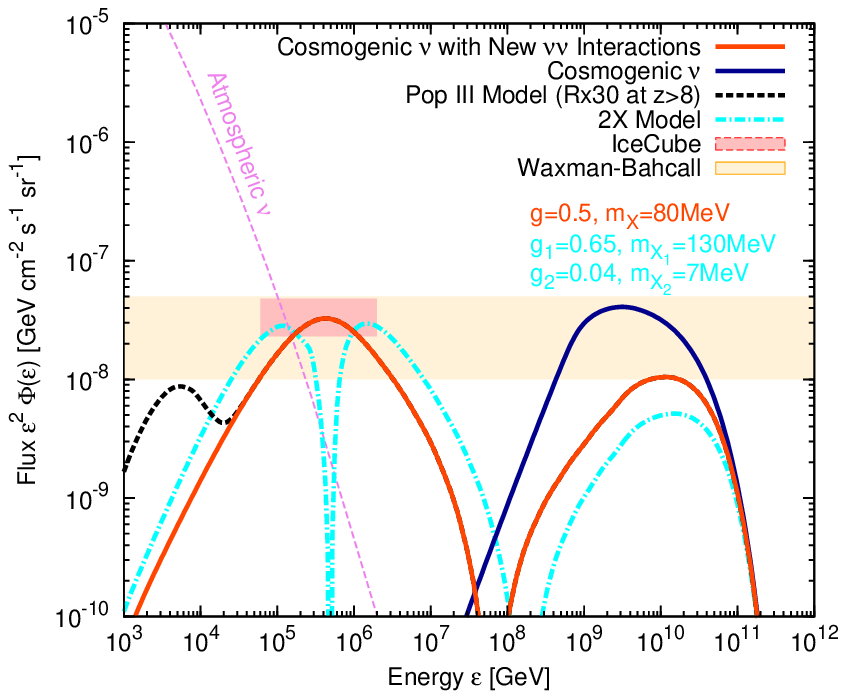}
  \end{center}
  \caption{
    The observed spectra of cosmogenic neutrinos
    with ({\it orange solid line}) and without ({\it blue solid line})
    the non-standard neutrino interactions
    in Eq.~(\ref{eq:signunu}) with $g=0.5$ and $m_X=80$ MeV.
    Also shown are the Pop III model with $30$ times higher
    rate $R(z)$ at $z>8$, and the 2$X$ model with two bosons,
    $g_1=0.65$, $m_{X_1}=130$ MeV, $g_2=0.04$, and $m_{X_2}=7$ MeV.
  }
  \label{fig:cascade}
\end{figure}

Figure~\ref{fig:cascade} shows the results of the observed flux
${\varepsilon_{\nu}^{\rm obs}}^2
\Phi_{\nu}=
{\varepsilon_{\nu}^{\rm obs}}^2
(c/4\pi) f_{\nu}(\varepsilon_{\nu}^{\rm obs},0)$
[GeV cm$^{-2}$ s$^{-1}$ sr$^{-1}$]
by calculating Eqs.~(\ref{eq:dfpdt})--(\ref{eq:zevo})
with an implicit method.
Within the Standard Model, the cosmogenic neutrino flux
peaks at $\varepsilon_{\nu} \sim 1$ EeV \citep[e.g.,][]{bz69,Yoshida_Teshima93,Takami+09}.
Invoking the hidden interactions in Eq.~(\ref{eq:signunu}) 
with $g=0.5$ and $m_{X}=80$ MeV,
the cosmogenic neutrinos initiate cascades with the C$\nu$B
and reappear at $\varepsilon_{\nu} \sim 1$ PeV 
where the universe becomes transparent
since the opacity peaks at $\varepsilon_{\nu} \sim m_{X}^2/2 m_{\nu}
\sim 10^{17}$ eV.
The spectral shape reflects the low-energy cross section
in Eq.~(\ref{eq:signunu})
and could be compatible with the IceCube data
given uncertainties of the atmospheric prompt neutrino background.


If the UHECR energy generation rate is high at high redshifts, 
e.g., from Population III star activities
\cite{Iocco+08,Suwa_Ioka11,de Souza+11},
we expect an enhancement at low energies, as 
in Fig.~\ref{fig:cascade} (Pop III model).
If there are two or more mediators, e.g., mu- and tau-types,
the spectrum may have dips
(2$X$ model).
A dip could lead to the possible deficit
in the IceCube data.

\section{Discussions}

Requiring that the observed neutrinos are not affected, 
we obtained astrophysical constraints on 
the secret interactions.
Although our work is greatly simplified,
detailed studies are possible,
including various interaction types,
scattering angles of particles,
and the neutrino--antineutrino difference.
Although a large coupling is already disfavored 
by the laboratory decay constraints,
an appealing point 
of the cascade scenario is that the PeV flux can 
coincide with the Waxman--Bahcall bound.
We must tune the mass to bring a peak to PeV energies,
but the neutrino flux is the same for different parameters.
The cascade scenario predicts the suppression of 
$>$PeV neutrinos,
whatever the source is.
Future neutrino detectors
such as
the Askaryan Radio Array \cite{ARA12} 
can test this possibility.

\ack
We thank J.~Beacom, K.~Blum, H.~Kodama, K.~Kohri, T.~Moroi, K.~Ng, Y.~Okada, K.~Omukai, and H.~Takami,
and also A.~Ishihara, K.~Mase, and S.~Yoshida for holding the workshop ``Cosmic Neutrino PeVatron (NuPeV 2014).''
This work is supported by
KAKENHI 24000004, 24103006, 26287051.\\
{\it Note added}: As this paper was being completed, we learned of an independent study by Ng and Beacom \cite{Ng+14}, which will be submitted to arXiv simultaneously.

\end{document}